\documentclass[pof, preprint,longbibliography,onecolumn]{revtex4-1}
\usepackage{textcomp}
\usepackage{amsmath}
\usepackage{amsfonts}
\usepackage{amssymb}
\usepackage{graphicx}
\usepackage{color}
\usepackage{epstopdf}

\begin{document}
\title{Brownian dynamics of elongated particles in a quasi-2D isotropic liquid}

\author{Christoph Klopp}
\affiliation{Otto-von-Guericke-Universit\"{a}t Magdeburg, Inst. for Experimental Physics, Dept. of Nonlinear Phenomena, 39016 Magdeburg, Germany}
\author{Ralf Stannarius}
\affiliation{Otto-von-Guericke-Universit\"{a}t Magdeburg, Inst. for Experimental Physics, Dept. of Nonlinear Phenomena, 39016 Magdeburg, Germany}
\author{Alexey Eremin}
\affiliation{Otto-von-Guericke-Universit\"{a}t Magdeburg, Inst. for Experimental Physics, Dept. of Nonlinear Phenomena, 39016 Magdeburg, Germany}


\begin{abstract}
We demonstrate experimentally that the long-range hydrodynamic interactions in an incompressible quasi 2D isotropic fluid result in an anisotropic viscous drag acting on elongated particles. The anisotropy of the drag is increasing with increasing ratio of the particle length to the hydrodynamic scale given by the Saffman-Delbr\"uck length. The micro-rheology data for translational and rotational drags collected over three orders of magnitude of the effective particle length demonstrate the validity of the current theoretical approaches to the hydrodynamics in restricted geometry. The results also demonstrate crossovers between the hydrodynamical regimes determined by the characteristic length scales.
\end{abstract}

\startpage{1}
\endpage{2}
\maketitle

\section{Introduction}

Motion of particles and their hydrodynamic interactions are of paramount interest not only for fundamental physics, but also for its applications in chemistry and biology. Liquid membranes are essential constituents of living matter. The mobility of particles (inclusions) embedded in biological membranes is known to determine many cellular processes such as signal transduction, stimuli response and sensing \cite{Simons:1997wy, RichardAStein:2002eb}. Self-assembly of membrane proteins can influence biochemical reactions and even allows cells to sense their shape \cite{Schweizer:2012ba}. Therefore, understanding of dynamics in membranes is of paramount interest. Membrane proteins and lipid rafts have often sizes significantly larger than the membrane thickness, and they can be viewed as macroscopic objects immersed in a continuous 2D fluid \cite{Simons:1997wy, Steffen:2001dn, RichardAStein:2002eb, Weiss:2013co}. Earlier studies of the dynamics of such structures revealed discrepancies in the determination of the membrane viscosities if the motion was described by 3D hydrodynamics \cite{Poo:1974ti,Hughes:2006vl,Gambin:2006jd}. For instance, an overestimation of the viscosity up to two orders of magnitude was found in experiments using fluorescent probes and NMR techniques \cite{Hughes:2006vl}.
Quantitative experimental data are rather scarce. Active microrheology studies of isotropic poly(dimethysiloxane) (PDMS) films on a fluid substrate revealed an appreciable discrepancy between experiment and theory \cite{Anguelouch:2006hf, Lee:2009ks}. This discrepancy was partially attributed to the compressibility of the PDMS layer. Here, we use smectic freely suspended films, which are unique models of nearly incompressible quasi-2D fluids allowing variation of hydrodynamic parameters in a very broad range. Both in-plane isotropic (smectic A) and in-plane anisotropic (smectic C) fluid structures can be explored in this system.

An inclusion moving with the velocity $\mathbf{v}$ experiences the viscous force $\mathbf{F}$. The viscous drag coefficients $\zeta_{\alpha\beta}$
are given by the inverse mobility tensor, defined by the expression:
\begin{equation}
	\zeta_{\alpha\beta}v_{\beta}=F_{\alpha}, \quad \mbox{with  }  \alpha, \beta = x, y, z.
\end{equation}

The viscous drag on a spherical inclusion in a viscous 3D fluid at low Reynolds number is proportional to the radius of the particle (3D Stokesian regime). However, the situation in a 2D fluid is more complex. In the case of an infinitely extended membrane in vacuum, the mobility is expected to diverge (Stokes-Paradox). But as demonstrated by Saffman and Delbr\"uck, the coupling between the flow in the membrane with the flow of the fluid or the fluids surrounding the membrane will result in a finite mobility, no matter how small the viscosities $\eta',\eta''$ of the outer fluids are \cite{Saffman:1976uh}.
The relation between the size of the inclusion and the drag force becomes a logarithmic function of the inclusion size. The coupling is described by a hydrodynamic length scale, the Saffman-Delbr\"uck length, $L_s=\eta_m/(\eta'+ \eta'')$ defined as the ratio between the 2D membrane viscosity $\eta_m$ and the 3D viscosities of the outer fluids $\eta',\eta''$ below and above the membrane, where $\eta_m=\eta h$ for a membrane with material viscosity $\eta$ and thickness $h$ \cite{Saffman:1976uh,Saffman:2003ve,Petrov:2008kg, Hughes:2006vl}. For an inclusion with characteristic size $L$, $L_s$ determines the Boussinesq number $\mathrm{Bo}=L_s/L$ which quantifies the relative contributions of the interfacial and bulk drag  \cite{Lee:2009ks}. When the lateral dimensions of the membrane reach $L_s$ or become smaller than  $L_s$, coupling to the membrane boundary dominates and determines the dynamics of inclusions. Altogether, three different dynamic regimes can be distinguished: the 3D Stokesian regime, the 2D regime driven by the coupling to the outer fluid and the 2D regime governed by the lateral spatial constraints of the membrane. All those regimes were demonstrated experimentally for in-plane isometric inclusions (circular cross-section) \cite{Nguyen:2010hu, Eremin:2011hk}. In lipid membranes, depending on the membrane viscosities, both 3D and 2D dynamics have been reported \cite{Cicuta:2007ci}. Yet, in some cases, it was demonstrated that the continuous approach breaks down for small proteins, where the inhomogeneities  of the membrane structure on a molecular scale may affect the protein diffusion  \cite{Gambin:2006jd}.

However, often the shape of the inclusions cannot be assumed  isometric and the problem of mobility of extended bodies in a 2D fluid must be considered in more detail \cite{Magusin:1993ey,Poo:1974ti, Gambin:2006jd, Stone:2015cs}. Saffman \cite{Saffman:1976uh} mentioned that the mobility of ellipses or ellipsoids can be easily calculated in a similar way as for discs. An example of this calculation is given in \cite{Stone:2015cs}. A rigid cylinder, perhaps one of the simplest anisometric shapes, exhibits
distinctive hydrodynamical properties even in  3D fluids. Already in the limit of low Reynolds number, the purely local character of the hydrodynamic drag is broken. Viscous drag on a rod is anisotropic and exhibits a logarithmic length-dependence:
\begin{equation}
	\zeta^{3D}_{\parallel}=2\pi\eta L\Big(\ln\Big(\frac{AL}{a}\Big)\Big)^{-1}
\end{equation}
\begin{equation}
	\zeta^{3D}_{\perp}=2 \zeta^{3D}_{\parallel}	
	\label{drag3D}
\end{equation}
where $\zeta^{3D}_{\parallel}$ and $\zeta^{3D}_{\parallel}$ are the drag coefficients for a motion parallel and perpendicular to the axis of the rod, respectively, $L$ is the rod length, $a$ is its radius, $\eta$ is the dynamic viscosity of the fluid, and $A$ is a numerical factor of the order of unity \cite{Doi:1994ug}. The topic of the present paper is an experimental study of the viscous drag and its dependence on the rod size in a 2D fluid. A theoretical model for the mobility of rod-shaped inclusions was put forward by Levine et al. \cite{Levine:2002jl}. They demonstrated that in contrast to the 3D case, the transverse drag coefficient $\zeta_{\perp}$ for larger rods becomes linear in length $L$, indicating that the viscous drag becomes purely local. In contrast, the parallel component $\zeta_{\parallel}$ of the drag for $L \gg L_s$ depends logarithmically on the length of the rod. At the same time, the relation in Eq.~(\ref{drag3D}) breaks down in 2D.
Rotational drag for rotation around the transversal axis in 3D is given by
\begin{equation}
  \zeta_{\rm{R}\perp}^{ 3D}=\frac{\pi\eta L^3}{3(\ln(L/2a)-C)}
\end{equation}
where $C$ is a numerical factor which depends on the aspect ratio of the rod and is approximately $0.662$ for infinitely thin rods \cite{Tirado:1984iv,Doi:1994ug}.

In this paper, we study the mobility of anisometric rod-shaped particles in a freely-suspended liquid crystal film by analysing their Brownian motion. Variation of the length of the rods and the thickness of the film enables us to explore a wide range of hydrodynamic regimes determined by the ratios of the typical rod dimensions to the Saffman-Delbr\"uck length $L_s$. 
We choose the commercial liquid crystal 4-n-octyl-4'-cyanobiphenyl (8CB) which exhibits the smectic-A phase at room temperature. Freely-suspended films with thicknesses from about 4 nm up to 5 \textmu{}m were used in our experiments. The preferential orthogonal alignment of the 3.17 nm long molecules without in-plane order at the film surfaces provides a unique opportunity to mimic a quasi-2D isotropic fluid. A sufficiently large extension of the film, in the range of a few centimeters, abates undesired boundary effects. Glass rods of various lengths (60 \textmu m to 120 \textmu m) width of 15 \textmu m were used as passive inclusions (Fig. \ref{MSD} a).

\section{Experimental}

\begin{figure}
 \centering
  \includegraphics[width=0.5\columnwidth]{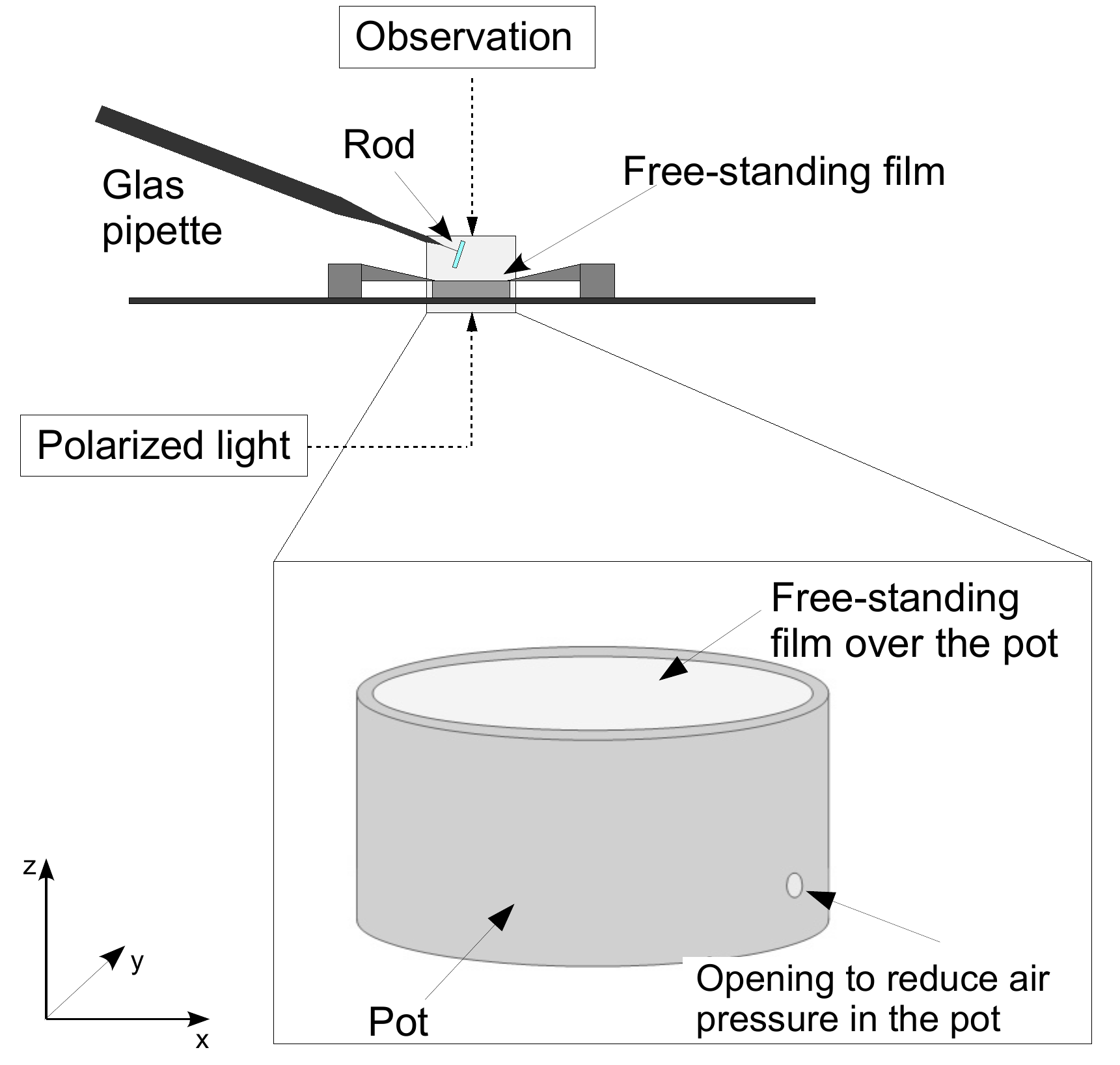}
\caption{Experimental setup for the study of Brownian motion in a freely suspended liquid crystal film. The film is drawn on an airtight sealed cylindrical container. Before deposition of the particles, we bend the smectic films lightly downward by an depression in the container, to avoid immediate drift of the particle into the meniscus. The pressure is then slowly equilibrated so that the experiment is performed with a flat film.
\label{setup}
}
\end{figure}

The commercial liquid crystal 4-n-octyl-4'-cyanobiphenyl (8CB) which exhibits the smectic-A phase at room temperature and forms freely-suspended films was used. The liquid crystal has a dynamic viscosity of $\eta=0.052$ Pa$\cdot$s at a room temperature of $T=22^{\circ}$C and is surrounded by ambient air with a viscosity of $\eta'=1.8 \cdot 10^{-5}$ Pa$\cdot$s. The observations were made using an AxioScope Pol polarising microscope (Zeiss GmbH). Glass rods of various lengths ranging from 60 \textmu m to 1500 \textmu m and widths of 15 \textmu m  were deposited on the film using a thin glass fibre attached to a micromanipulator (Fig. \ref{setup}). The relative measurement uncertainties of translation and rotational drag were estimated to be below 10\%.

\section{Results}
\begin{figure}
\centering
  \includegraphics[width=\columnwidth]{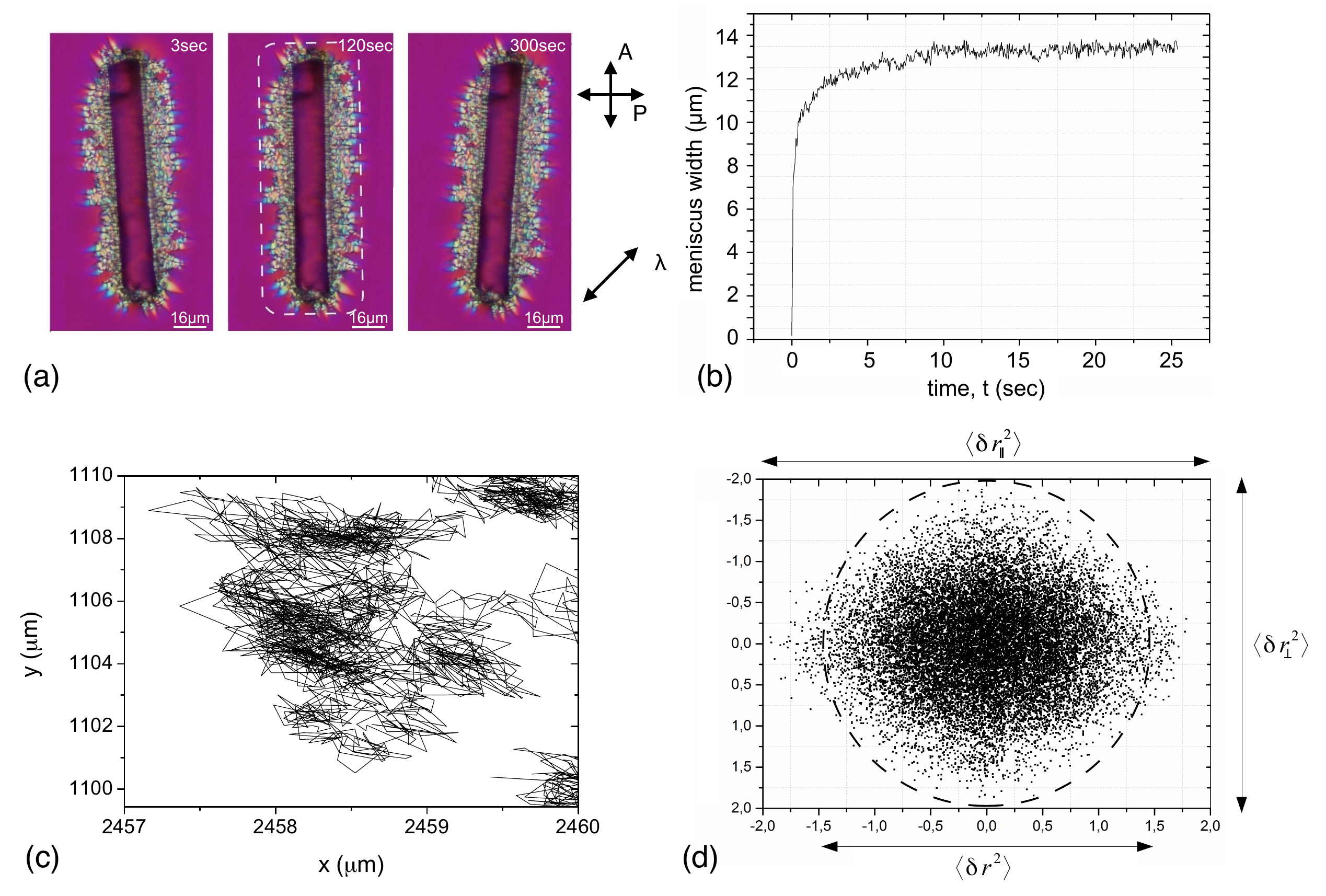}
\caption{(a) Polarising microscopy images of a rod embedded in the freely suspended film, observed with a full wave plate. The images are taken at time instances of 3, 120 and 300 s after the deposition. The bright birefringent area of the meniscus indicates a complex inner structure where the smectic layers are deformed. Its boundary schematically marked by a dash line. The arrows on the right show the orientations of the polariser P, analyser A, and the waveplate. (b) Time dependence of the meniscus width for the rod in (a). (c) A typical trajectory of a rod and (d) a distribution of displacements measured between positions on
the trajectory separated by the time interval of 1/60 s. \label{MSD}
}
\end{figure}

After deposition on the film, the glass rod is rapidly wetted by the LC material. A meniscus within about 3 - 5 sec (Fig. \ref{MSD}(a,b)). Polarising microscopy reveals focal-conic textures in the meniscus  (Fig. \ref{MSD}(a)), which indicates the presence of complex internal structures composed by layer dislocations and topological defects. For this reason, the flow in the thick meniscus area cannot be considered as purely two-dimensional. It is substantially impeded by the internal structure of the meniscus, evidenced by the stationary optical textures of the decoration pattern. Consequently we have to consider the meniscus area as immobile respective to the rod, and we introduce an effective dimension of the rod, by correcting it with the size of the meniscus:
\begin{equation}
  L=L_{\rm{rod}} +2L_{\rm{m}},
\quad
  W=W_{\rm{rod}} +2W_{\rm{m}}
\end{equation}
where $L_{\rm{rod}}$ and $W_{\rm{rod}}$ are the length and the width of the bare glass rod, $L_{\rm{m}}$ and $W_{\rm{m}}$ are mean lengths and widths of the meniscus, respectively.

\begin{figure}
 \centering
  \includegraphics[width=\columnwidth]{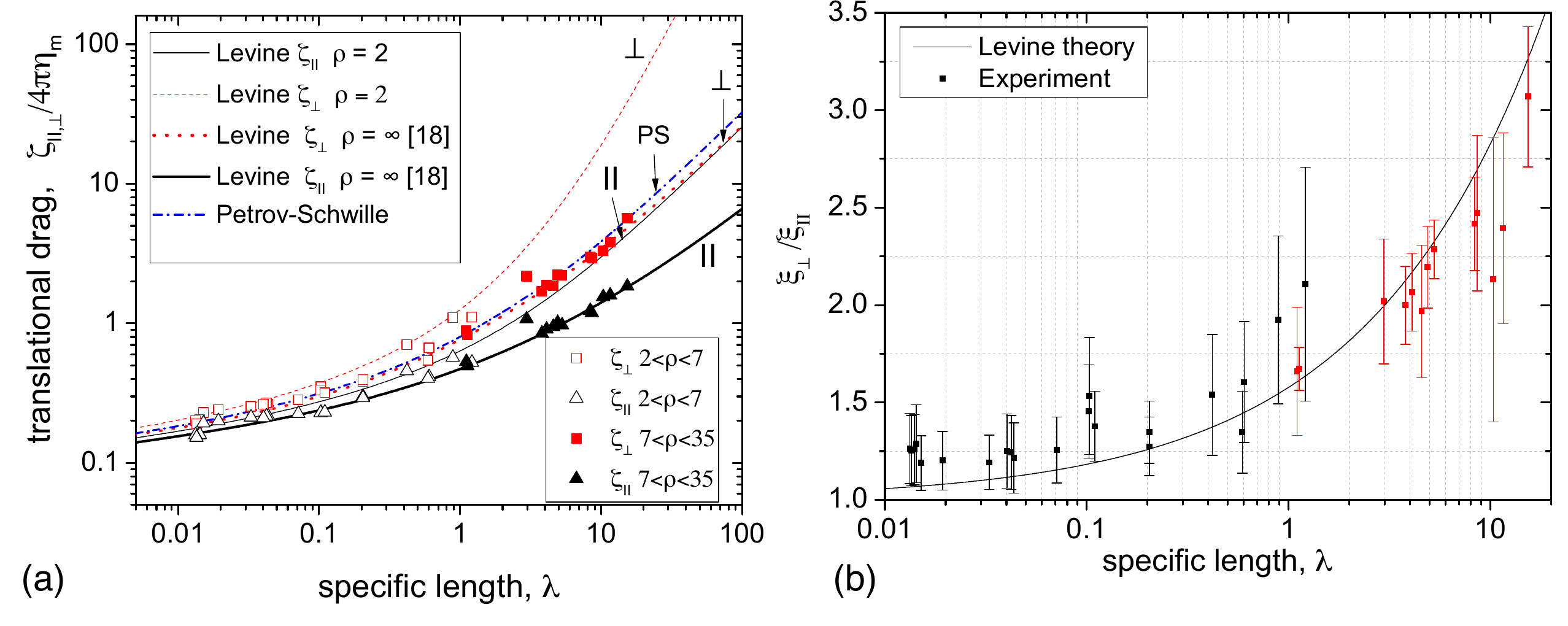}
\caption{(colour online) (a) Parallel and perpendicular drag coefficients for thin rods as a function of their dimensionless length $\lambda=L/L_s$. The solid lines are theoretical predictions by the Levine model \cite{Levine:2004jj} for a rod aspect ratio $\rho=2$ and for infinitely thin rods ($\rho=\infty$). The dash-dotted line corresponds to the Petrov-Schwille equation with an argument $\lambda=L/L_s$ \cite{Petrov:2008kg}. The experimental data points are separated in two groups of rods with different aspect ratios $\rho=L/W$: $2<\rho<7$ (open symbols) and $7<\rho<35$ (filled symbols). (b) Ratio of the perpendicular component of the viscous drag to the parallel component, $\zeta_{\perp}/\zeta_{\parallel}$ as a function of $\lambda$. The solid line is the theoretical prediction for thin rods from Ref.~\cite{Levine:2004jj}.
\label{drag}
}
\end{figure}

The Brownian motion of the rods was recorded over time intervals of 5 to 6 min, an example of such a trajectory is shown in Fig.~\ref{MSD}(c). In order to determine the viscous drag $\zeta$ acting on the rod in a freely suspended film, we separate the Brownian motion from some unavoidable drift which impedes the measurement at long time scales. The separation of the two motions is achieved with a procedure described earlier   \cite{Nguyen:2010hu}. The transversal and longitudinal components of the drag are obtained from the mean squared displacement (MSD) relative to the rod axes, by separating the displacements parallel to the rod's momentary long  and short axes (Fig.~\ref{drag}(d)).

The experimental results for the translational  drag coefficients in dependence on the effective rod length are shown in Fig. \ref{drag}. Drag coefficients for different film thicknesses and rod lengths are scaled to the reduced rod length $\lambda=L/L_s$. The width of the rods including the meniscus is smaller than $L_s$ in all cases.  A first obvious result is that the diffusion along the rod
axis, as expected, is faster than perpendicular to it. This feature is more pronounced when $\lambda>1$. For $\lambda\approx 10$, a factor of about 2.5 is reached. The translational drag exhibits a distinct non-linear behaviour as a function of the reduced rod length. Approximating the rods by discs with the effective radius  $R_{\rm{eff}}=L/2$ (the mean radius $R_{\rm{m}}=(L+W)/4 \rightarrow L/2$ for $W \rightarrow L$), and setting $\lambda=2R_{\rm{eff}}/L_s$,
we can see that the experimental data for the mean translational drag cannot be satisfactorily approximated by the 2D drag model proposed by Petrov and Schwille (PS) as an extension of the Saffman-Delbr\"uck theory for isometric particles (discs) \cite{Petrov:2008kg, Saffman:1976uh, Saffman:2003ve}. The mean drag is systmatically lower than the drag predicted by the PS model. Only for  $\lambda\ll 1$, the experimental data are close to the PS prediction (Fig. \ref{drag}(a)). It is important to note that the PS drag in Fig. \ref{drag}(a) is plotted as a function of $L/L_s$ In the case $\lambda\ll 1$, the aspect ratios $\rho=L/W$  of the rods used in our study are not large, and the mean radius is close to $L/2$.
For larger $\lambda$, the drag coefficients for the displacements parallel to the rod axis, $\zeta_{\parallel}$, and perpendicular to it, $\zeta_{\perp}$, diverge from each other and the displacements become strongly correlated with the momentary rod axis. The transversal drag $\zeta_{\perp}$ becomes a linear function of the length $\lambda$. The parallel drag $\zeta_{\parallel}$ shows a slower continuous growth with $\lambda$.

\section*{Discussion}
In order to calculate the drag on the anisometric particles in 2D, we followed the theoretical approach developed by 
Levine et al. \cite{Levine:2004jj,Levine:2004kd,Levine:2002jl}, where velocity response functions parallel 
($\chi_{\parallel}$) and perpendicular ($\chi_{\perp}$) to the rod long axis were used.
Specific dimensionless drag coefficients $\zeta_{\parallel, \perp}/4\pi \eta_m$ were computed as functions of the reduced length $\lambda$ using the Kirkwood approximations in Fig. \ref{drag}(a).
Two pairs of curves are included in Fig.~\ref{drag}(a): the theoretical dependences $\zeta_{\parallel, \perp}/4\pi \eta_m$ for the rod aspect ratio of 2, computed using the Kirkwood approximation \cite{Doi:1994ug, Levine:2004kd},
and the thin-rod approximation from Ref.~\cite{Levine:2004jj}. The fifth curve is the drag on a disc-shaped inclusion 
with radius $L/2$, computed using the Petrov-Schwille equation \cite{Petrov:2008kg}. 
This curve covers both, the 2D Saffman-Delbr\"uck regime for small $\lambda$ and 3D Stokes regime for $\lambda\gg 1$.

The smallest value of $L_s$ achieved in our experiment is about 40 \textmu m, which is  comparable with the widths of the rods used in our experiments ($W \approx 40 - 45$ \textmu m). However, for thicker films, $L_s$ significantly exceeds the width $W$. This justifies using the theoretical model for thin rods. In agreement with the theory by Levine et al.~\cite{Levine:2004kd}, even for thick rods, the translational drag coefficients become independent of $\rho$, and can be described in the thin-rod approximation, when $\lambda$ is sufficiently small  (Fig.~\ref{drag}(a)).
 The plot in Fig. \ref{drag}(a) shows an excellent agreement between the experimental drag and the theory for the 2D hydrodynamics. It is remarkable  that no fitting parameters are used here. The anisotropy of the translational diffusion is already measurable for the shortest rods with the specific lengths, $\lambda$, as low as 0.01, and it increases with increasing length (Fig.~\ref{drag}(b)). Even when $\lambda\ll 1$, the anisotropy of the viscous drag remains appreciable. The two curves for the transversal and longitudinal drag coefficients converge logarithmically. In the limit $\lambda \rightarrow 0$, the rod represents a singular distortion of the flow field and the structure of the rod on a scale much smaller than the Saffman-Delbr\"uck length $L_s$ becomes less important.
   The drag coefficients become nearly independent of the inclusion size and approach the values given by the PS formula \cite{Petrov:2008kg} for isometric particles with a diameter of $L$ in an isotropic membrane. This can be understood from the asymptotic behaviour of the response functions, $\chi_{\parallel}$ and $\chi_{\perp}$, which leads to the Saffman-Delbr\"uck drag force in the limit of small $\lambda$.


\begin{figure}
 \centering
  \includegraphics[width=0.45\columnwidth]{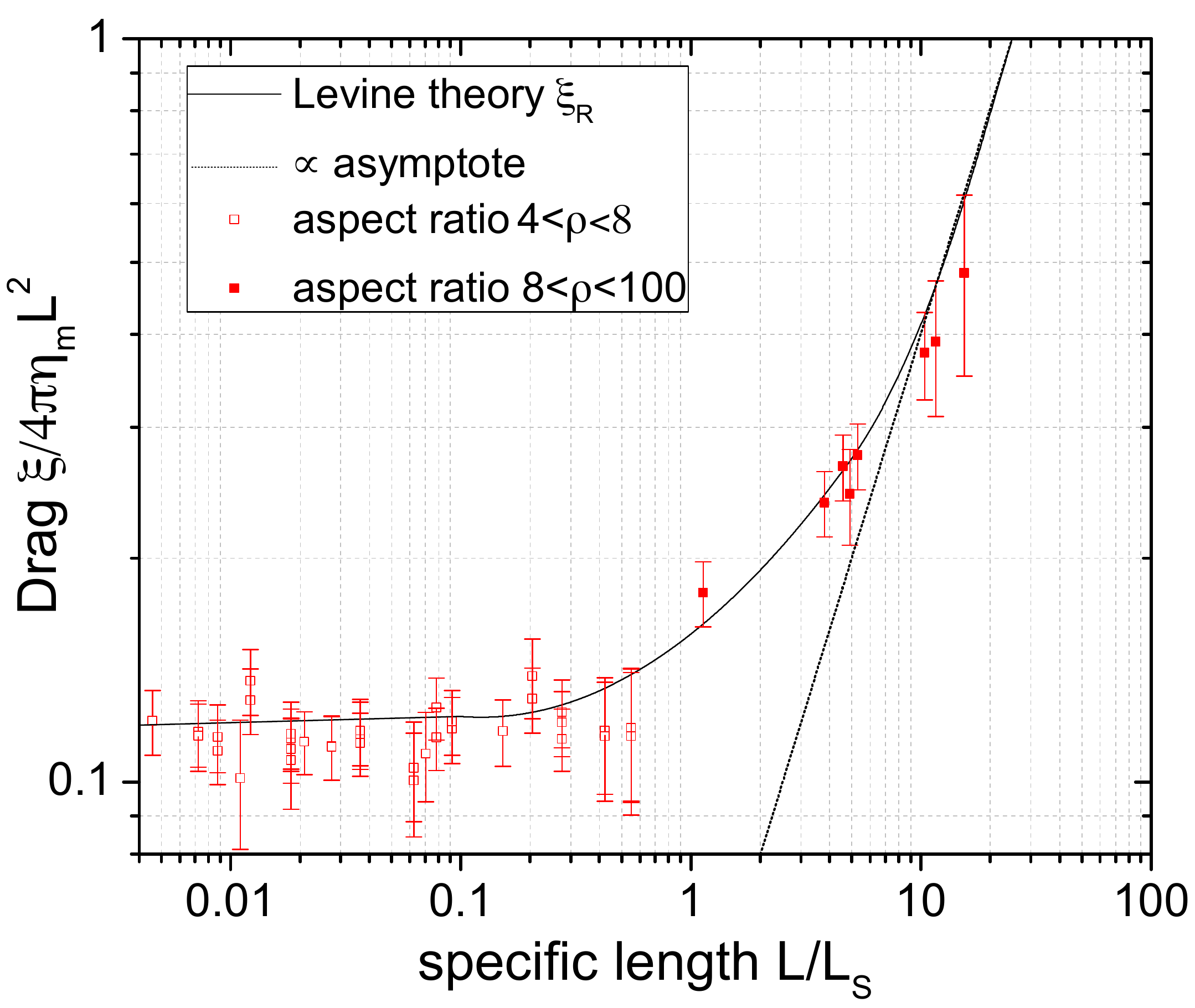}
\caption{Dependence of the rotational drag coefficient $\zeta_R$ on the specific length $\lambda$. The experimental data are compared with the prediction of the Levine theory \cite{Levine:2004jj} (solid line) for infinitely thin rods.
\label{rot_dif}
}
\end{figure}

The rotational drag coefficient weighted by the length squared, $\zeta_R/4\pi\eta_m L^2$, is shown in Fig. \ref{rot_dif}. In agreement with the theoretical prediction by Levine et al. \cite{Levine:2004jj,Levine:2004kd,Levine:2002jl}, the drag coefficient exhibits an algebraic dependence on the rod length in the asymptotic limits. It is proportional to $\lambda^2$ for $\lambda\ll 1$ and proportional to $\lambda^3$ for $\lambda \gg 1$. Our experimental data fully cover the quadratic regime of $\zeta_R$ and show the cross over to the third-power regime for $\lambda> 1$. Remarkably, these results are different from those obtained using active microrheology in oil (PDMS) films \cite{Lee:2009ks}, where the agreement with the Levine's theory was significantly poorer, especially in the range of small $\lambda$. This discrepancy was attributed to the compressibility of the oil layer on a fluid substrate and the flow of the oil over the inclusion. In the case of smectic films, however, the film is nearly incompressible and a change of the film thickness in response to a compressive stress may occur only through the nucleation of dislocations. The latter can be excluded in our experiment. 

The meniscus around the inclusion is composed of a complex layer structure featuring regular arrangements of topological defects and dislocations on a microscopic scale. This inhibits flow within the decorated regions around the inclusions. At the same time, the film outside the meniscus is perfectly uniform and flat on the hydrodynamic scale, so that there is no influence of potential curvatures on the viscous dynamics \cite{Camley:2012cq, Morris:2015jj}.

In summary, we reported an experimental study of translational and rotational viscous drag coefficients for rod-shaped inclusions in an isotropic quasi-2D fluid, modeled by a freely-suspended liquid crystal film. The drag coefficients show a distinct nonlinear dependence on the rod size. The translational drag exhibits an anisotropic behaviour already for the smallest aspect ratios and is in an excellent agreement with the theory by Levine et al. \cite{Levine:2004jj}. 
For inclusion thicker than the film, an immobile meniscus has to be accounted for in the effective size of the inclusion. 
We confirm a crossover behaviour of the translational drag coefficient, $\zeta_{\perp}$, to a regime where the drag is purely local and depends linearly on the length $L$. For sufficiently small lengths $L\ll L_s$, the translational drag coefficients can be well described by the thin-rod approximation. 

Notably, the {\em transversal} mobility of thin rods can be reasonably well described by the Petrov-Schwille equation, where the radius of an equivalent circular inclusion is taken as $L/2$. The enormous aspect ratio of freely suspended smectic films in combination to the easy adjustment of homogeneous
film thicknesses from nanometers to micrometers allowed quantitative measurements of the diffusion of rods over more than three orders
of magnitude in the effective rod length $\lambda$, which is hardly achievable with any other physical system.

%
%
%
%
%
%


\subsection{Acknowledgement}
The authors acknowledge funding by DFG within project STA 425/28-1 and C. K. acknowledges a Landesstipendium Sachsen-Anhalt. J.E. Maclennan (University of Colorado) is acknowledged for fruitful discussions.

\bibliography{papers}

\end{document}